\documentclass[12 pt,reqno]{revtex4}
\usepackage{epsfig}
\usepackage{amssymb}
\newcommand{\be}{\begin{equation}}
\newcommand{\ee}{\end{equation}}
\newcommand{\ba}{\begin{eqnarray}}
\newcommand{\ea}{\end{eqnarray}}

\begin{document}

\title{The optimized Rayleigh-Ritz scheme for determining the quantum-mechanical spectrum}

\author{Przemys\l aw Ko\'scik and Anna Okopi\'nska\\
Institute of Physics, \'Swi\c{e}tokrzyska Academy\\
 \'Swi\c{e}tokrzyska 15, 25-406 Kielce, Poland}


\begin{abstract}

\noindent The convergence of the Rayleigh-Ritz method with nonlinear
parameters optimized through minimization of the trace of the
truncated matrix is demonstrated by a comparison with analytically
known eigenstates of various quasi-solvable systems. We show that
the basis of the harmonic oscillator eigenfunctions with optimized
frequency $\Omega$ enables determination of bound-state energies of
one-dimensional oscillators to an arbitrary accuracy, even in the
case of highly anharmonic multi-well potentials. The same is true in
the spherically symmetric case of $V(r)=\frac{\omega^2
r^2}{2}+\lambda r^k$, if $k>0$. For spiked oscillators with $k<-1$,
the basis of the pseudoharmonic oscillator eigenfunctions with two
parameters $\Omega$ and $\gamma$ is more suitable, and optimization
of the later appears crucial for a precise determination of the
spectrum.
\end{abstract}

\maketitle
\section{Introduction}
Since the early days of quantum mechanics, the techniques based on
the variational principle have been successfully used for
determination of ground-state characteristics of various physical
systems. In its simplest form, the variational approach utilizes a
single trial function that depends upon certain parameters, the
values of which are fixed so as to minimize the expectation value of
the Hamiltonian. This approach enables a very accurate determination
of the ground-state energy, if the functional form of the trial
function is appropriately chosen, but its extension to higher
bound-states appears impractical because of difficulties in assuring
orthogonality of the trial functions. A more suitable approach is
offered by the \textit{linear Rayleigh-Ritz} (RR) method with a
trial function represented as a finite linear combination \be
\phi(\overrightarrow{r})=\sum_{n=0}^{N-1}{c_{n}}\psi_{n}(\overrightarrow{r}),\ee
where the functions $\psi_{n}(\overrightarrow{r})$ are taken from
some orthonormal basis in the function space. Treating the
coefficients $c_{n}$ as variational parameters, one obtains a set of
linear equations
\be\sum_{n=0}^{N-1}(H_{mn}-\varepsilon\delta_{mn})c_{n}=0,m=0,1,...,
N-1\label{sec}\ee where $H_{mn}=<m|\widehat{H}|n>$ denotes the
matrix element of the Hamiltonian between the states of the basis,
$|n>=\psi_{n}$. The solution of the above equations yields $N$th
order approximations to wave functions
$\phi_{i}^{(N)}(\overrightarrow{r})$, and the corresponding energies
$\varepsilon_{i}^{(N)}$, for \mbox{ $i=0,...,N-1$} states. The
accuracy of the RR method can be systematically improved by
increasing the number $N$ of basis functions, obtaining successive
approximations to the larger and larger number of states. In this
way a desired part of the spectrum may be determined with the
approximate eigenvalues monotonically converging to the exact
bound-state energies~\cite{RS} at a rate strongly dependent on the
choice of the basis. The method is also known under the name of
\textit{exact diagonalization} although the results become exact
only in the limit as $N$ tends to infinity. Long time ago, Hylleraas
observed~\cite{Hyl} that the effectiveness of the RR method may be
further improved by introducing a scale parameter into the functions
of the basis. Since then, various sets of functions depending on
various nonlinear parameters have been used in determining spectra
of atoms and molecules. The orthonormal sets made up of the
eigenfunctions of a solvable Hamiltonian which depends on certain
free parameters are especially convenient. In chemical applications
the values of nonlinear parameters are fixed so as to a optimize the
convergence of the RR estimate to the ground state energy; in the
early works, the best values were found by trial and
error~\cite{kolos} at present, they are fixed in computationally
demanding procedures of iterative
optimization~\cite{rych,AOsingular}.

Optimization of the RR scheme may performed on the basis of the
principle of minimal sensitivity (PMS), which has been successfully
used to improve perturbative calculation~\cite{PMS,cash,killPT}. PMS
requires the values of unphysical parameters introduced into
calculation to be chosen so as to make approximations to physical
quantities as less sensitive to the variation of these parameters as
possible. This suggest to improve the RR method by fixing the values
of nonlinear parameters so as to minimize the $N$th order
approximation to the desired level energy $E_{n}^{N}$. Such a
strategy requires however diagonalization of the RR matrix to be
performed in an algebraic way, which is feasible but only in low
orders~\cite{AOsingular}, or the application of extensive procedures
of iterative minimization. Moreover, the scheme is not very
economical, since the whole procedure must be repeated for each
considered state. The optimized RR scheme, proposed by one of
us~\cite{ao}, adopts a different strategy that is also based on the
PMS but insists on fixing the values of nonlinear parameters before
diagonalization of the truncated matrix. Since the only physical
quantity that can be determined before diagonalization of the $N$th
order RR matrix is its trace \be
Tr_{N}H=\sum_{n=0}^{N-1}<n|\widehat{H}|n>, \label{trace}\ee which
represents the sum of $N$ bound-state energies, we require the
values of nonlinear parameters to be chosen so as to render
$Tr_{N}H$ stationary. The advantage of the scheme is that the $N$th
order approximations to many eigenstates are determined in one run
and the obtained approximation to wave functions are orthogonal. It
has been shown~\cite{ao} that a good accuracy is acquired for the
quartic anharmonic oscillator, using a basis of the harmonic
oscillator (HO) eigenfunctions with optimized frequency. In this
work we show that also in the case of other interaction potentials,
the stationarity of the $Tr_{N}H$ condition optimizes the choice of
nonlinear parameters. We note that implementation of the method with
a modern software environment allows arbitrary precision
calculations, which is especially important for determining the tiny
energy splitting in the case of multi-well potentials with nearly
degenerate levels. The computational cost of the optimized RR scheme
is not high, as the 50-digits accuracy of lowest bound-state
energies is easily achieved with RR matrices of order $N<100$, even
in the most difficult cases of multi-well oscillators. Through the
example of a sextic oscillator with analytically known eigenstates
we show that the performance of the method for wave functions is
also good, as various moments of the position operator appear well
convergent. Later, we discuss the case of spherically symmetric
potentials, where the two-parameter set of pseudo-harmonic
oscillator (PHO) eigenfunctions constitutes a more appropriate basis
for the optimized RR method. We gauge the convergence of the method
for various anharmonic potentials $\lambda r^k$, of both positive
and negative power $k$, comparing the results with the exact
solutions of different quasi-solvable examples (the sextic
oscillator, the harmonium system, and the spiked oscillator). In all
the cases studied, a good convergence is automatically ensured by
using the trace condition for fixing the nonlinear parameters.

The plan of our work is as follows. In section~\ref{HO}, we study
the optimized RR method for one-dimensional anharmonic oscillators
using the basis of the HO eigenfunctions. In section~\ref{spher},
the basis of the PHO eigenfunctions is introduced and the
performance of the method for spherically symmetric potentials is
discussed. Section~\ref{con} is devoted to conclusion.
\section{One-dimensional case}\label{HO}
\subsection{Harmonic oscillator basis}\label{HOmethod}
The success of the RR method depends on the appropriate choice of
the basis in the functional space. A basis constructed from the HO
eigenfunctions \be
\psi_{n}^{\Omega}(x)=\left(\frac{\Omega}{\sqrt{\pi}2^{n}n!}\right)^{\frac{1}{2}}H_{n}(\Omega
x)e^{-\frac{\Omega^2x^2}{2}}\label{HObasis}\ee with an arbitrary
frequency $\Omega$ playing a role of a nonlinear parameter has
proved convenient for solving one-dimensional problems with purely
discrete
spectrum~\cite{McW,balsa,jafar,extfield,bishop,QM,chaudhuri,kill,korsch,hill}.
We use the above basis for determining spectrum of various
anharmonic oscillators (AO).
\subsection{Quartic oscillator}\label{Quartic}
The most popular example is the quartic AO with the Hamiltonian
operator given by \be
\widehat{H}=-\frac{1}{2}\frac{d^2}{dx^2}+\frac{\omega^2
x^2}{2}+\lambda x^4,\label{AO}\ee where the units $\hbar=1$ and
$m=1$ are used. Bound-states of the system, tq
hat exist if
$\lambda\geq 0$, correspond to those solutions of the
Schr\"{o}dinger equation \be \widehat{H}\phi(x)=E\phi(x)\ee which
vanish at infinity. Mc Weeny and Coulson noted~\cite{McW} that the
HO wave functions (\ref{HObasis}) with any value $\Omega
>0$ constitute an appropriate basis in the RR
calculations for the quartic AO, as they fulfill the bound-state
condition, but the convergence of the scheme depends strongly on the
value of $\Omega$. They observed that the approximation series for
the $n$th state energy converges quickly if the value of $\Omega$ is
fixed so as to minimize the matrix element $<n|H|n>$. However, such
a prescription  works well only in the case of a single-well AO
($\omega^2>0$) but fails if the potential is double-well shaped
($\omega^2<0$) ~\cite{balsa}. Moreover, diagonalization procedure
has to be repeated with a different value of $\Omega$ for each
considered state. A more effective strategy is to take
approximations to all the desired states from diagonalization of a
single RR matrix with a compromise value of $\Omega$. Many ways of
fixing the value of $\Omega$ have been tested: so as to minimize the
expectation value of the Hamiltonian in a chosen state of the HO
basis (the first, namely $|0>$ in~\cite{jafar}, the central
in~\cite{extfield}, the last, namely $|N>$ in~\cite{bishop}, that
one for which the expectation of the Hamiltonian is the smallest
in~\cite{balsa}), so as to minimize a sum of expectation values in
several HO states~\cite{QM}, or so as to vanish the matrix element
$<2|H|0>$~\cite{chaudhuri}. However, none of these prescriptions may
be justified by the PMS, as none of the considered quantities does
represent an approximation to a physical quantity. In this
connection, one of us proposed~\cite{ao} to fix the nonlinear
parameters so as to make the trace of the truncated matrix
stationary, since $Tr_{N}H$ represents the $N$th order approximation
to the sum of energies of the $N$ lowest bound-states. For the HO
basis~(\ref{HObasis}), this amounts to fixing the frequency in $N$th
order calculation to the value $\Omega_{opt}^{(N)}$, which fulfils
\be \frac{d}{d\Omega}Tr_{N}H=0.\label{opty}\ee As shown in
Ref.~\cite{ao}, the scheme automatically yields well-converging
results for the spectrum of the quartic AO, in both the single- and
double-well cases. We may add here that the values
$\Omega^{(N)}_{opt}$, determined by the solution of (\ref{opty}),
appear close to those for which the convergence of the above RR
scheme is the quickest. This is demonstrated in Table~\ref{kuku},
where $\Omega^{(N)}_{opt}$ are compared with the values
$\Omega^{(N)}_{min}$, obtained by a numerical minimization of the
error for the ground-state energy. A choice of nonlinear parameters
has been considered on the same example of a quartic oscillator from
a different perspective in Ref.~\cite{hill}, where an analytic
formula for optimum value has been derived from the asymptotic
expansion. Comparing Table VIII of Ref.~\cite{hill} (where $\alpha$
corresponds to our $\Omega^2$) with our Table~\ref{kuku}, we may see
that the values derived from the asymptotic optimization scheme
appear closer to $\Omega^{(N)}_{min}$ than our values
$\Omega^{(N)}_{opt}$, but an extension of this scheme to other
systems and other basis sets is problematic, since asymptotic
expansions which are uniformly valid in the nonlinear parameters are
difficult to construct. Whereas, a fully algorithmic formulation of
our optimization scheme allows an easy application for arbitrary
systems. Generally, the accuracy of the optimized RR method
diminishes with increasing number of the level, but the energies of
$N/2$ lowest states can be trusted on, and using their values for
calculating the free energy of the system, provides highly accurate
results in a broad temperature range~\cite{ao}. Similarly, the
well-determined part of the spectrum has been successfully utilized
for approximate description of the time evolution in the quartic
oscillator potential~\cite{hugh}.
\begin{table}[h]
  \centering
  \begin{tabular}{|c|c|c|c|c|}
    \hline
    N & $\sqrt{\Omega^{(N)}_{opt}}$& $\delta E_{0}^{(N)}(\Omega_{opt})$ &
    $\sqrt{\Omega^{(N)}_{min}}$& $\delta E_{0}^{(N)}(\Omega_{min})$ \\
    \hline
    1  & 1.29294233500847 & 1.09716442402927 $10^{-2}$ & 1.29294233500847 & 1.09716442402927 $10^{-2}$\\
5 & 1.65920419620602 &
  8.63597923540916 $10^{-7}$& 1.70670645005687 & 1.33517756262170 $10^{-7}$\\

10 & 1.86080663733626 & 1.86447406929386 $10^{-12}$& 1.95004181438340 & 1.69012176051751 $10^{-13}$\\

15 & 1.98940338014799 & 6.71884044353837 $10^{-19}$ & 2.09841072018140 & 2.7410652496 $10^{-19}$\\

20  & 2.08593508969090 & 4.36194667514212 $10^{-24}$ & 2.20749144840388 & 4.9101 $10^{-25}$\\
    \hline
  \end{tabular}
  \caption{The optimum value of the nonlinear parameter $\Omega^{(N)}_{opt}$
  determined from the trace condition and the corresponding relative error of
  the ground state energy, $\delta E_{0}^{(N)opt}(\Omega_{opt})$, compared with
  $\Omega^{(N)}_{min}$, taken from Ref.~\cite{hill}, for which the relative error of ground-state energy is
  minimal, $\delta E_{0}^{(N)}(\Omega_{min})$.} \label{kuku}
\end{table}

It is worthwhile noting that the optimized RR scheme may be
implemented in a way allowing an arbitrary precision calculation by
taking advantage of the present computer algebra abilities to deal
with exact numbers. As an example we use the Mathematica package to
calculate energy difference $\Delta E$ between the first excited and
the ground state of the double-well oscillator with a Hamiltonian $
\widehat{H}=-\frac{g}{2}\frac{\partial^2}{\partial
x^2}+\frac{1}{2g}(x^2-\frac{1}{4})^2$ for $g=.001$. Diagonalisation
of the RR matrix of order $N=350$ with the Mathematica precision
constant set to $250$, yields the value of the level splitting \ba
\Delta
E=1.470464454175092501381989964494151981567800350052603\nonumber
\\52838605333378036605041575193505284182673433993282124674887\nonumber\\312560803946942066500267
7938176074660629119~10^{-68},\ea which agrees with the 225 digits
quoted with the result obtained on the basis of Zinn-Justin
conjecture and confirmed by high-precision power series
method~\cite{AL}.
\subsection{Sextic oscillator bound-states}\label{HOnum}
The optimized RR scheme performs also well for oscillators with
higher than quartic anharmonicities, which present a more stringent
test for approximation methods. We consider the sextic AO with a
Hamiltonian given by \be
\widehat{H}=-\frac{1}{2}\frac{d^2}{dx^2}+{\omega^2\over{2}}
x^2+\lambda x^6\label{gen}\ee that provides an interesting
quasi-solvable example~\cite{CM}. This is one of the rare cases when
several bound-states of a system may be obtained exactly, namely
$p+1$ eigenstates are known if the parameters of the sextic
oscillator satisfy the condition \be
\omega^2=-(3+4p+2\nu)\sqrt{2\lambda},\label{condsex}\ee where
$\nu=0$ corresponds to the even-parity eigenstates and $\nu=1$ to
the odd-parity ones. The detailed explanation of how to obtain the
exact eigenvalues and the explicit form of the corresponding
wave-functions for a chosen value of $p$ is given in the Appendix.

\begin{figure}[h]
\begin{center}
\includegraphics[width=0.8\textwidth]{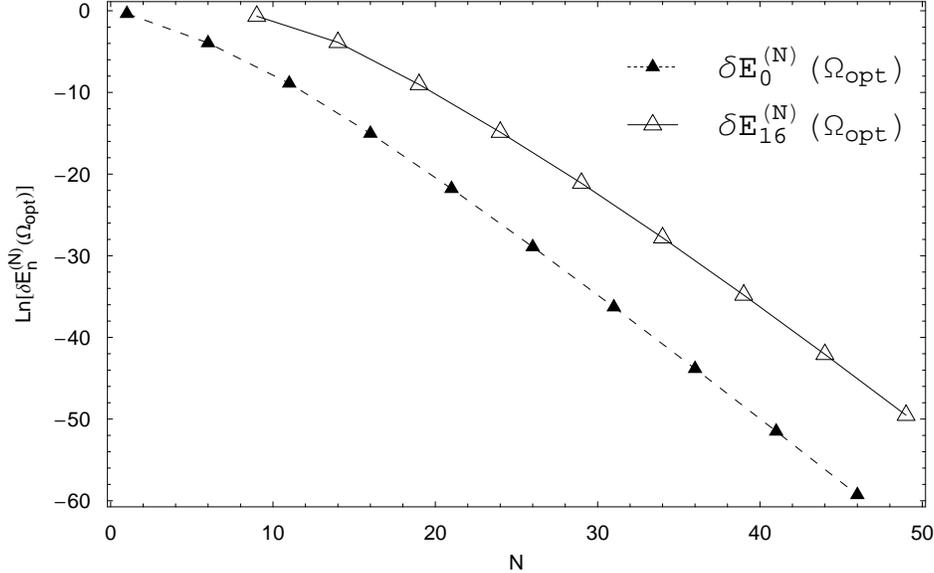}
\end{center}
 \caption{Semilogarithmic plot of
the relative energy error, $\delta E_{n}^{(N)}(\Omega_{opt})$, for
n=0 and n=16 state of the sextic oscillator (\ref{geneven}) in
function of the dimension $N$ of the optimized RR matrix. Here and
in the following figures lines are drawn to guide the eye.}
\label{onedimensionoscillator}
\end{figure}

Bound-state energies $E(\omega,\lambda)$ depend on the parameters of
the potential~(\ref{gen}), however in calculating the numerical
results, we may set $\lambda=1$ without loss of generality. The
eigenenergies at other values of $\lambda$ can be obtained as
multiples of $E(\omega\lambda^{-\frac{1}{4}},1)$, according to the
scaling relation $E(\omega,\lambda)=\lambda^{\frac{1}{4}} E(
\omega\lambda^{-\frac{1}{4}},1),$ obtained by applying the
transformation $x\mapsto\lambda^{-\frac{1}{8}} x$ to the
Schr\"{o}dinger equation. As an example for testing the convergence
of the optimized RR scheme we choose here the even parity case with
$p=8$, which according to (\ref{condsex}) corresponds to the
quasi-solvable Hamiltonian \be
\widehat{H}|_{\nu=0}=-\frac{1}{2}\frac{d^2}{dx^2}-\frac{35\sqrt{2}}{2}
x^2+ x^6\label{geneven}\ee with a family of nine exactly known
states ($n=0,2,...,16$). The $N$th approximation to the AO
bound-states, $|n\gg$, is obtained by numerical diagonalization of
the Hamiltonian matrix in the basis of the N lowest even-parity wave
functions of the HO~(\ref{HObasis}) with a frequency
$\Omega^{(N)}_{opt}$ determined from~(\ref{opty}). In Table
\ref{even0one} and \ref{even16one} the approximate values of energy
and various moments of the position operator $
x_{n}^{k(N)}(\Omega_{opt})=\ll n|x^k|n\gg$ are given for the
ground-state $(n=0)$ and for the highest of the analytically known
excited states $(n=16)$, respectively. A quick convergence of all
the moments in Tables \ref{even0one} and \ref{even16one} indicates
that wave functions are accurately determined. We have checked
indeed that the agreement between the approximate and exact wave
functions to graphical accuracy is obtained already at $N=15$. An
exponential convergence for bound-state energies is evidenced in
Fig.\ref{onedimensionoscillator}, where the relative error $\delta
E_{n}^{(N)}(\Omega_{opt})$ is plotted as function of $N$. In the
case of the Hamiltonian $
\widehat{H}|_{\nu=1}=-\frac{1}{2}\frac{d^2}{dx^2}-\frac{37\sqrt{2}}{2}
x^2+ x^6$, when the family of nine odd-parity states is exactly
known, the convergence properties of the optimized RR method are
similarly good,  but they slowly worsen with increasing value of
$\omega$, when the double-well potential becomes deeper and more
bound-states are known exactly.

Also in case of higher anharmonicities, the RR method optimized by
the trace condition enables highly accurate determination of the
spectrum in relatively low order calculation. For instance,
diagonalisation of matrices of order $N<100$ is sufficient for
reobtaining all the bound states of the most difficult multi-well AO
examples collected in~\cite{near-exact} with the precision gradually
diminishing from 50-digits for the ground state to 40-digits for the
tenth excited state. The accuracy of various moments of the position
operator is also good, as the hypervirial relations are fulfilled to
a very high precision. Instead of presenting extensive tables of
results, we refer the reader to our web page with the Mathematica
program~\cite{web} which may be easily adapted for solving the
desired AO problem to the desired precision.

\begin{table}[h]
  \centering
  \begin{tabular}{|c|c|c|c|c|}
    \hline
    N & $ E_{0}^{(N)}(\Omega_{opt})$ & $x_{0}^{2(N)}(\Omega_{opt})$ & $x_{0}^{6(N)}(\Omega_{opt})$& $x_{0}^{10(N)}(\Omega_{opt})$    \\
    \hline
 20  &  -\underline{40.5262528}0948697&    \underline{2.7264316}6988877& \underline{23.6063074}7666730& \underline{242.23182}746777999\\
 25  &  -\underline{40.5262528234}2447& \underline{2.7264316738}8909& \underline{23.6063074825}4632& \underline{242.2318224}2437541\\
30  &  -\underline{40.5262528234356}6&  \underline{2.7264316738926}8& \underline{23.60630748253}900& \underline{242.2318224178}8116\\
35  &  -\underline{40.52625282343567}&    \underline{2.72643167389269}&\underline{23.60630748253899}&\underline{242.23182241787521}\\

    \hline
  \end{tabular}
   \caption{The approximate energies $E_{0}^{(N)}(\Omega_{opt})$
and moments of the position operator $x_{0}^{k(N)}(\Omega_{opt})$
for \mbox{$k=2,6,10$}, obtained by means of the optimized RR method
for the ground-state ($n=0$) of the sextic oscillator
(\ref{geneven}) at specific values of the dimension $N$. The
underlined digits agree with the exact results.}\label{even0one}
\label{min}
\end{table}

\begin{table}[h]
  \centering
  \begin{tabular}{|c|c|c|c|c|}
    \hline
    N  & $E_{16}^{(N)}(\Omega_{opt})$&  $x_{16}^{2(N)}(\Omega_{opt})$ & $x_{16}^{6(N)}(\Omega_{opt})$& $x_{16}^{10(N)}(\Omega_{opt})$    \\
    \hline
 20  &  \underline{40.52}790329985854 &  \underline{2.037}02842810644&  \underline{35.34}043326592744 & \underline{856}.72311216403564\\
 25  & \underline{40.52625}631425536&    \underline{2.03737}443716647 &\underline{35.342}77775827479&\underline{856.421}83863495585\\
30  &\underline{40.52625282}743947 &   \underline{2.03737569}324205&\underline{35.3428011}2924913&\underline{856.421252}73736842\\
35  &  \underline{40.52625282343}881  &\underline{2.03737569518}445&\underline{35.34280117889}126&\underline{856.42125207}650909\\
40  &  \underline{40.52625282343567}&  \underline{2.03737569518631}&\underline{35.342801178949}60&\underline{856.421252075969}73\\

    \hline
  \end{tabular}
  \caption{Same as Table \ref{even0one}, but for state $n=16$.}\label{even16one} \label{min}
\end{table}

\section{Spherically symmetric case}\label{spher}
The optimized RR method can be easily extended to the case of a
central potential $V(r)$, where $r=|\overrightarrow{r}|$, when the
Schr\"{o}dinger equation reduces to the one-variable problem at
fixed angular momentum $l$. In the three-dimensional space, the
problem is represented by the equation \be[-{1\over 2r}{d^2\over
dr^2}r+{l(l+1)\over 2r^2}+V(r)]R(r) =ER(r),\label{sphR}\ee which
upon introducing $u(r)=r R(r)$ is transformed to the form \be
\widehat{H}^{(l)}u(r) =Eu(r)\label{Ssph}\ee with the radial
Hamiltonian \be \widehat{H}^{(l)}=-{1\over 2 }{d^2\over
dr^2}+{l(l+1)\over 2r^2}+V(r),\label{Hsph}\ee and boundary
conditions $u(r)\rightarrow 0$, as $r \rightarrow \infty$ and
$u(0)=0$.

\subsection{Pseudoharmonic oscillator basis}\label{PHO}
q The
completeness of the set has been shown by Hall at al.~\cite{spiked}.
For $A=l(l+1)$, i.e. $\gamma=l+\frac{3}{2}$, the
solution~(\ref{Lag}) reduces to that of a spherically symmetric HO
with a Hamiltonian \be \widehat{H}_{HO}=-{1\over 2}{d^2\over
dr^2}+{\omega^2 r^2\over 2}+{l(l+1)\over 2r^2}\label{ho}.\ee

Optimization of nonlinear parameters of the PHO basis through the
trace condition (\ref{trace}) amounts to choosing the values of
$\Omega$ and $\gamma$ in $N$th order RR calculation so as to satisfy
\be
\frac{d}{d\Omega}Tr_{N}H^{(l)}=0,~~\mbox{and}~~\frac{d}{d\gamma}Tr_{N}H^{(l)}=0.\label{opty2}\ee

\subsection{Radial oscillators}
Among interesting examples that can be easily treated by the
optimized RR method are the radial oscillators described by the
Hamiltonian of the form \be \widehat{H}^{(l)}=-{1\over 2r}{d^2\over
dr^2}r+{l(l+1)\over 2r^2}+{\omega^2 r^2\over 2}+\lambda
r^{k}\label{anh}\ee with powers of anharmonicity $k$ being both
positive and negative. The above hermitian symmetric Hamiltonian is
semi-bounded and consequently has self-adjoint extension in the
Hilbert space $L_{2}(0,\infty)$ which admits a spectral
decomposition~\cite{Fun}. Its spectrum is purely discrete. The
Schr\"{o}dinger equation for the above Hamiltonian may be rescaled
in two ways: the transformation $r\mapsto  r
\lambda^{\frac{1}{(k+2)}}$ leads to the relation
$E(\omega,\lambda)=\lambda^{\frac{2}{(k+2)}} E(z,1)$ and the
transformation $r\mapsto
 r \omega^{-\frac{1}{2}}$ yields $E(\omega,\lambda)=\omega E( 1,z^{\frac{-(k+2)}{2}})$
 with the dimensionless parameter $z=\omega \lambda^{-\frac{2}{(k+2)}}$. This
shows that power $k=-2$ sets the border between very different
behavior of the oscillators. Denoting by $\kappa$ a positive value
$|\frac{2}{(k+2)}|$, in the case of $k>-2$ we have $z=\omega
\lambda^{-\kappa}$ and $E(\omega,\lambda)=\lambda^{\kappa} E(z,1)$
and $E(\omega,\lambda)=\omega E( 1,z^{-\kappa})$, which shows that
for $\lambda\rightarrow 0$ or $\omega\rightarrow 0$, the weak
coupling or strong coupling perturbation theory may be respectively
applied. The cases of $k<-2$ (spiked oscillators) are very
different. For spiked oscillators, $z=\omega \lambda^{\kappa}$,
$E(\omega,\lambda)=\lambda^{\kappa} E(z,1)$ and
$E(\omega,\lambda)=\omega E( 1,z^{\kappa}),$ which indicates that
neither of the two terms of the interaction potential may be taken
as dominant and the conventional perturbation expansions cannot be
applied. The Klauder phenomenon~\cite{Klauder} does not occur for
radial problems with the boundary conditions of the Dirichlet type,
i.e. $u=0$ at the singular point $r=0$, and for $\lambda\rightarrow
0$ the energy eigenvalues of a spiked oscillator converge to those
of the spherically symmetric HO of frequency $\omega$~(\ref{ho}).
However, the perturbation series contains the terms logarithmic in
$\lambda$ and is ordered in fractional powers of $\lambda$, thus
unconventional methods have to be invoked for its
derivation~\cite{Klauder,Harrell}. The perturbation theory with
respect to $\widehat{H}_{HO}$ is supersingular because the matrix
elements $\langle m|r^k|n\rangle$ in the HO basis are infinite.
Similar difficulties would be encountered for spiked oscillators in
conventional RR calculations with the HO basis. As pointed out by
Hall et al.~\cite{Basis}, the problem may be avoided by using the
PHO basis with $\gamma> -k/2$, since in this case the matrix
elements appear finite and the RR calculation are well defined. For
the anharmonic oscillators~(\ref{anh}) the numerical calculations
may be simplified by removing the dependence on $\Omega$ from the
matrix elements by rescaling $r\mapsto{r\over \sqrt{\Omega}},$ which
yields
\begin{eqnarray} H_{mn}^{(l)}=(2n+\gamma)\Omega\delta_{mn}+{(\omega^2-\Omega^2)\over 2\Omega}\langle
m|r^2|n\rangle-\nonumber\\{\Omega\over 2}(\gamma^2-2\gamma+{3\over
4}-l(l+1)) \langle m|{1\over r^2}|n\rangle +
{\lambda\over\Omega^{k/2}}\langle
m|r^k|n\rangle,\label{HPHOanh}\end{eqnarray} where the ket
$|n\rangle$ corresponds to the radial function (\ref{Lag}) with
$\Omega=1$, namely \be u_{n}^{\gamma}(r)=(-1)^{n}{1\over
\Gamma(\gamma)}\sqrt{2 \Gamma(\gamma+n)\over n!} r^{\gamma-{1\over
2}}e^{-{1\over 2} r^2}{_{1}F_{1}}(-n;\gamma;r^2).\label{Lag1}\ee

In this work we consider three different examples: the spherically
symmetric sextic oscillator ($k=6$), the harmonium potential
($k=-1$) and the "antisextic" spiked oscillator ($k=-6$), which all
enjoy the nice feature of quasi-exact solvability. We use the
explicitly known solutions for testing the performance of the
optimized RR method for both regular and singular spherically
symmetric oscillators. In the case of positive power anharmonicities
($k>0$), the optimum values of $\gamma$ turn out to be around
$l+3/2$, we may put therefore $\gamma=l+3/2$ and use the basis of
the radial HO eigenfunctions with frequency $\Omega$ being the only
parameter to be optimized. On the other hand, in the case of
singular anharmonic oscillators ($k<-1$), the role of the parameter
$\gamma$ is crucial, and we show that the values of $\gamma_{opt}$
determined from the trace optimization \label{opty2} are always
greater than $-k/2$, which ensures a successful calculation.

\subsubsection{Radial sextic oscillator\label{sphAO}}

\begin{figure}[h]
\begin{center}
\includegraphics[width=0.8\textwidth]{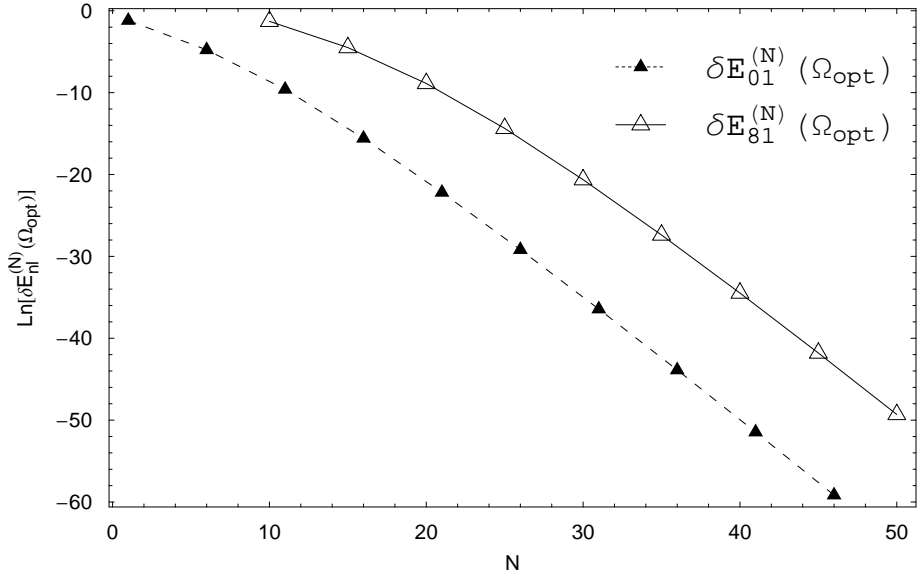}
\end{center}
 \caption{Semilogarithmic plot of
the relative energy error, $\delta E_{n1}^{(N)}(\Omega_{opt})$ for
the two states (n=0,8) of the radial sextic oscillator
(\ref{radial6}) as function of the dimension $N$ of the optimized RR
matrix.} \label{sphericaloscillator}
\end{figure}

First, we consider a sextic AO with the radial Hamiltonian of the
form \be\widehat{H}^{(l)}=-{1\over 2}{d^2\over dr^2}+{l(l+1)\over
2r^2}+{\omega^2\over 2} r^2+{\lambda r^6}.\label{rad6}\ee  As shown
in the Appendix, the wave functions of the $p+1$ lowest states are
known in a closed-form if the parameters of the above oscillator are
related as  \be \omega^2=-(5+4p+2l)\sqrt{2\lambda},\label{cond1}\ee
which is similar to that for the odd-partity  one-dimensional case
(\ref{condsex}) but includes in addition the orbital number $l$.
Besides a normalization factor $\frac{1}{\sqrt{2}}$, the s-wave
sector in a central potential is equivalent to the odd sector in a
one-dimensional potential, therefore we need only consider the case
$l>0$. Setting $\lambda=1$, we discuss the case of $p=8$, when \be
\widehat{H}^{(l)}=-{1\over 2}{d^2\over dr^2}+{l(l+1)\over
2r^2}+{(37+2l)\sqrt{2}\over 2} r^2+{ r^6},\label{radial6}\ee
comparing the nine analytically determined eigenstates with the
approximations obtained from diagonalization of the RR matrix in the
basis of the radial HO eigenfunction with optimized frequency
$\Omega$. The Tables \ref{energies0} and \ref{energies8} contain our
results for orbital number $l=1$ for the lowest $(n=0)$ and highest
($n=8$) of exactly known states. It can be observed how the
approximations to bound-state energies,
$E_{n1}^{(N)}(\Omega_{opt})$, and the moments
$r_{n1}^{k(N)}(\Omega_{opt})=<<n|r^k|n>>$ for $k=2,6,10$, approach
the exact values with increasing $N$. In
Fig.\ref{sphericaloscillator} the dependence of the relative energy
error on $N$ is shown for both states.

\begin{table}
  \centering
  \begin{tabular}{|c|c|c|c|c|}
    \hline
    N & $ E_{0}^{(N)}(\Omega_{opt})$ &  $r_{0}^{2(N)}(\Omega_{opt})$&  $r_{0}^{6(N)}(\Omega_{opt})$&  $r_{0}^{10(N)}(\Omega_{opt})$  \\
    \hline
 20  &  -\underline{48.1135341}7791531 &  \underline{2.90220193}255671&  \underline{27.988866}43153595&  \underline{314.77282}893841708\\
25  &   -\underline{48.1135341890}4174  &  \underline{2.90220193690}299&  \underline{27.988866516}87285&  \underline{314.77282672}623041\\
30  &   -\underline{48.11353418905196}& \underline{2.9022019369073}7&  \underline{27.988866516955}77&  \underline{314.77282672244}758\\
35  &   -\underline{48.11353418905196} &\underline{2.90220193690738}&  \underline{27.98886651695584}&  \underline{314.7728267224433}7\\
    \hline
  \end{tabular}
  \caption{The approximate energy $E_{0}^{(N)}(\Omega_{opt})$
and the moments of the radius operator $r_{0}^{k(N)}(\Omega_{opt})$
(\mbox{$k=2,6,10$}) determined by the optimized RR method for the
ground-state ($n=0$) of the radial sextic oscillator~(\ref{radial6})
for $l=1$. }\label{energies0} \label{min}
\end{table}

\begin{table}
  \centering
  \begin{tabular}{|c|c|c|c|c|}
    \hline
    N & $E_{8}^{(N)}(\Omega_{opt})$ & $r_{8}^{2(N)}(\Omega_{opt})$& $r_{8}^{6(N)}(\Omega_{opt})$& $r_{8}^{10(N)}(\Omega_{opt})$ \\
    \hline
 20   &\underline{48.11}737731862358&  \underline{2.175}21762647996&  \underline{42.02}668954998873&  \underline{113}6.40959750355092\\
25  &\underline{48.1135}4441726266&  \underline{2.1759}2836963907&  \underline{42.0313}2546920644&  \underline{1135.6}7058214554528\\
30   &\underline{48.113534}20292910& \underline{2.1759318}3660339&  \underline{42.031398}51074727&  \underline{1135.66913}552605609\\
35   &\underline{48.1135341890}6437& \underline  {2.1759318431}1651&  \underline{42.03139869962}153&  \underline{1135.66913419}959634\\
40  & \underline{48.1135341890519}7  & \underline{2.17593184312375}&  \underline{42.031398699877}43&  \underline{1135.669134198948}09\\
    \hline
  \end{tabular}
   \caption{Same as Table \ref{energies0}, but for the state $n=8$.}\label{energies8}
\label{min}
\end{table}

\subsubsection{Harmonium\label{harm}}

\begin{figure}[h]
\begin{center}
\includegraphics[width=0.8\textwidth]{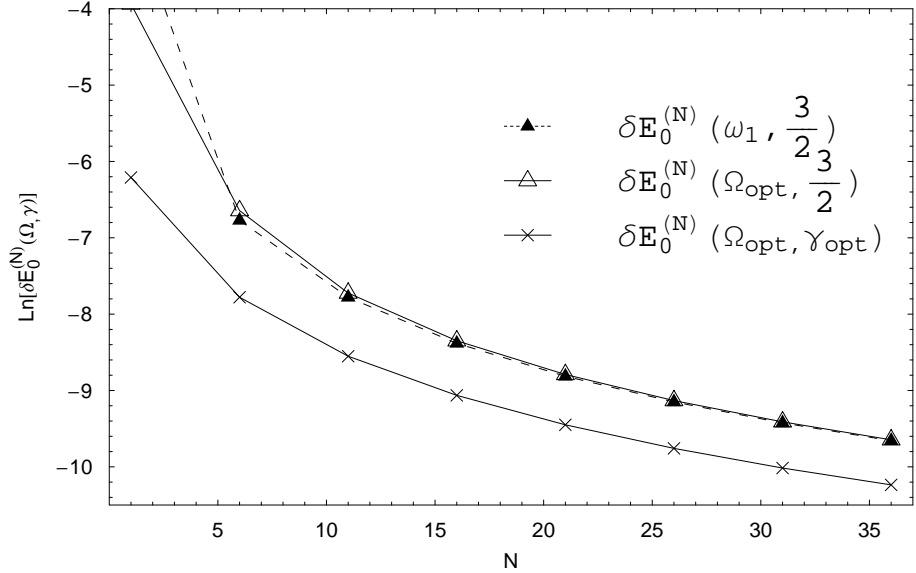}
\end{center}
 \caption{Semilogarithmic plot of the
relative energy error for the ground state of harmonium with
confinement frequency $\omega_{1}=0.25$ as function of the dimension
$N$ of the optimized RR matrix.} \label{harmonium025}
\end{figure}

\begin{figure}[h]
\begin{center}
\includegraphics[width=0.8\textwidth]{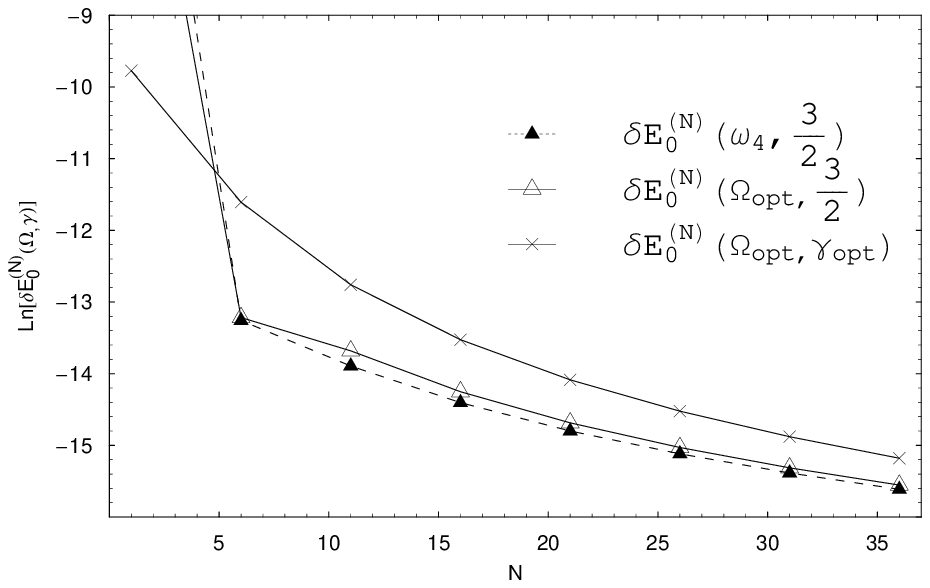}
\end{center}
 \caption{Like Fig.\ref{harmonium025}, but for confinement frequency $\omega_{4}\approx
0.00867$.}
 \label{harmonium4}
\end{figure}

The next example is related to the harmonium problem. Harmonium is a
system of two particles confined in a harmonic potential and
interacting via a Coulomb force, which enjoys the pleasant feature
that the center of mass and the relative motion can be separated.
The center of motion is subjected to a solvable harmonic oscillator
equation, and the spherically symmetric relative motion equation
corresponds to the $k=-1$ power AO with a radial Hamiltonian of the
form \be \widehat{H}^{(l)}=-{d^2\over dr^2}+{l(l+1)\over
r^2}+{\omega^2 r^2}+{\lambda \over r}.\label{harg}\ee The
observation that the above Hamiltonian possess a closed-form
solution~\cite{hersch,taut} was of great importance and thus
provides further rationale for investigation of approximation
methods in the many-body theory. The exact solution exists if for
some integer $p$, the frequency $\omega$ fulfils the condition \be
E=(3 + 2 l + 2 p)\omega, \label{condHar}\ee which, in contrast
to~(\ref{cond1}), does not depend on the coupling $\lambda$ but on
the energy of the investigated state. In difference with the case of
the sextic oscillator, for harmonium there is but a single
bound-state, not necessarily the ground-state, that is known
exactly. This happens if the parameters $\omega$ and $\lambda$ are
related by a particular relation that depends on an integer $p$, as
given in the Appendix~(\ref{Det}). By utilizing the scaling property
$E(\omega,\lambda)=\lambda^{2} E( \omega\lambda^{-2},1)$ and setting
$\lambda=1$ we have that Eq.~(\ref{Det}) determines the values
$\omega_{p}$ at which the bound-state is analytically known. Using
the exact bound-state solutions for testing the convergence of the
optimized RR method in the PHO basis, we compare three ways of
fixing the nonlinear parameters: first, a naive RR scheme
($\Omega=\omega_{p}$ and $\gamma=3/2+l$), second, optimization of
the parameter $\Omega$ ($\Omega=\Omega_{opt}$ and $\gamma=3/2+l$),
and third, optimization of both parameters ($\Omega=\Omega_{opt}$
and $\gamma=\gamma_{opt}$). Fig.\ref{harmonium025} presents the
semilogarithmic plot of the relative energy error $\delta
E_{nl}^{(N)}$ as function of $N$ for the largest possible value of
confining frequency for which the ground state solution is known,
namely $\omega_{1}=0.25~(l=0,p=1)$. A similar plot for a smaller
value of frequency, $\omega_{4}={35-3\sqrt{57}\over 1424}\approx
0.00867$ ($l=0,p=4)$, is given in Fig.\ref{harmonium4}. The results
for higher values of angular momentum $l$ show a similar tendency,
therefore we do not show them. In the case of large confinement
frequency the optimized scheme proves superior over the naive one,
but in the strong correlation limit ($\omega<<1$) it is the naive
one that works slightly better. We may notice that the application
of the trace condition in the RR scheme ensures an exponential
convergence, albeit its rate is generally slower than in the case of
positive power AOs. Since the values of nonlinear parameters
determined from the trace condition appear close to $\gamma=3/2+l$
and $\Omega=\omega_{p}$, the optimized RR scheme may be regarded as
a justification for using the naive RR method for harmonium-like
systems.

\subsubsection{Spiked oscillator}\label{singular}

Now, we come to the computationally more difficult case of a spiked
oscillator with anharmonic potential of the power $k<-2$, which
exhibits a highly singular behavior at the origin. As an example we
consider the "antisextic" oscillator ($k=-6$) with the radial
Hamiltonian given by \be \widehat{H}^{(l)}=-{1\over 2}{d^2\over
dr^2}+{l(l+1)\over 2r^2}+{\omega^2\over 2} r^2+{\lambda \over
r^6}.\label{sing}\ee A bound state of the system may be considered
analytically determined if its energy is given by \be
E^{(p)}=2(p+1)\omega\label{en}\ee and a specific relation between
$\omega$ and $\lambda$ (\ref{matrspiked}) is satisfied for an
integer $p$. As opposed to the previously discussed sextic
oscillator and harmonium cases, where the results have been rescaled
in terms of $\lambda$, in the case of spiked oscillators we exploit
the rescaling in terms of $\omega$, namely $E(\omega,\lambda)=\omega
E( 1,\omega^{2}\lambda)$. This allows us to put $\omega=1$ in the
numerical calculation in accordance with the usual practice in the
literature on spiked oscillators. The specific values of $\lambda$
at which the bound states are analytically known may be obtained by
substituting (\ref{en}) into (\ref{matrspiked}) and determining the
$p+1$ solutions  of the polynomial equation (\ref{Det}). If
$\lambda_{p(i)}$ are numbered with decreasing value of $\lambda$
from $i=0$ to $i=p$, then the $i$th exactly known state represents
the $i$th excitation in the respective potential.

\begin{figure}[h]
\begin{center}
\includegraphics[width=0.8\textwidth]{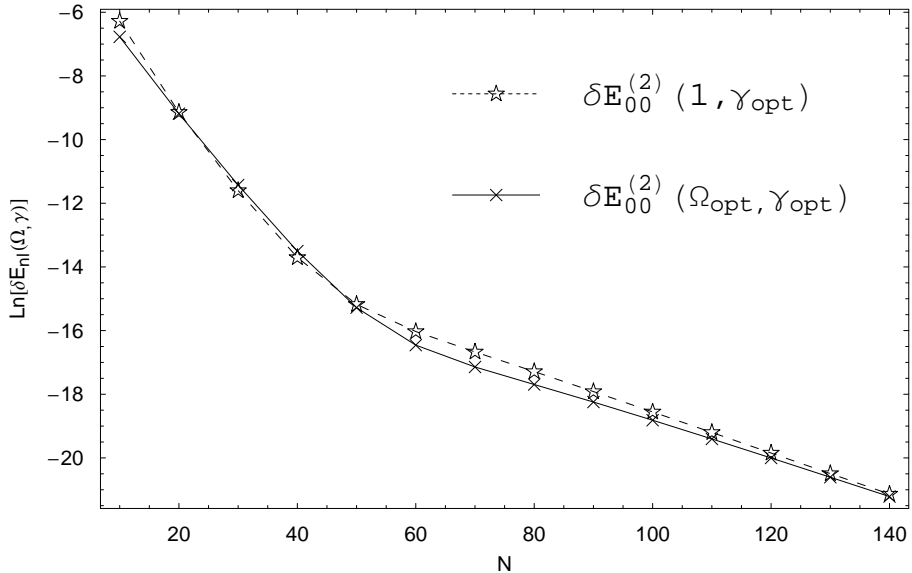}
\end{center}
 \caption{Semilogarithmic plot of
the relative energy error for the ground-state of the radial spiked
oscillator (\ref{sing})  of the strength $\lambda_{2(0)} \approx
369.26$ as function of the dimension $N$ of the optimized RR
matrix.} \label{singulareven2}
\end{figure}

\begin{figure}[h]
\begin{center}
\includegraphics[width=0.8\textwidth]{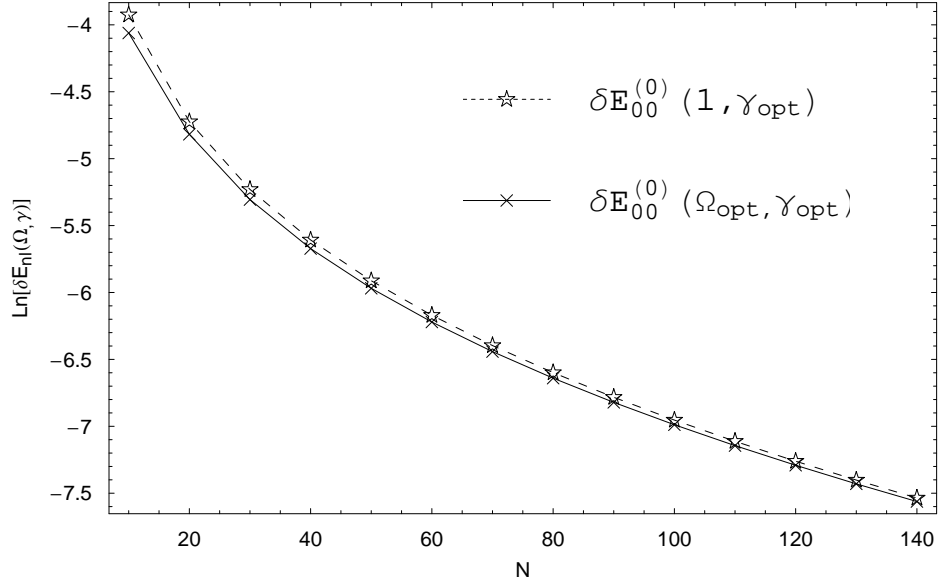}
\end{center}
 \caption{Like Fig.\ref{singulareven2}, but for the strength $\lambda_{0(0)}\approx 0.07$.}
\label{singulareven1}
\end{figure}

\begin{figure}[h]
\begin{center}
\includegraphics[width=0.8\textwidth]{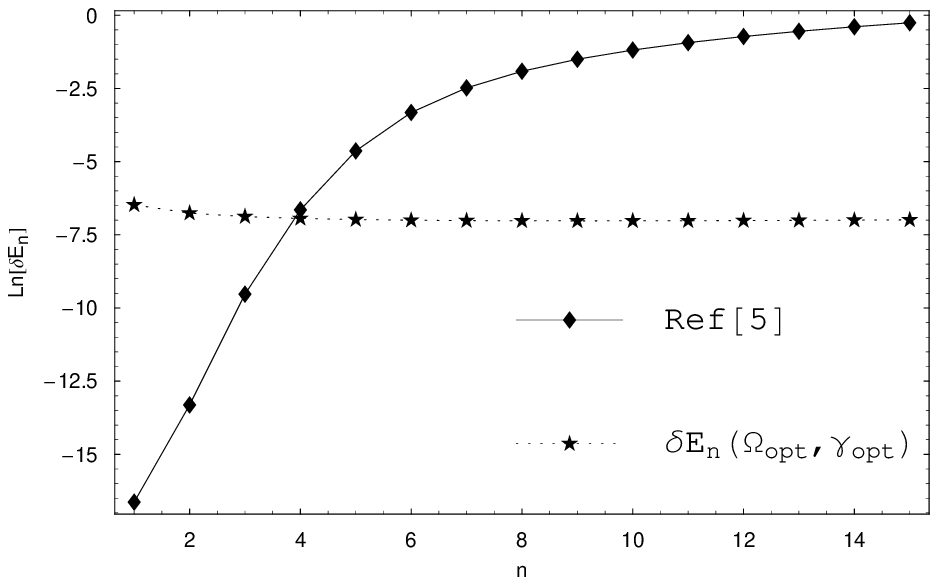}
\end{center}
 \caption{The semilogarithmic plot of the relative error of bound state
 energy of the spiked oscillator of $\lambda=0.1$
 as function of the level number $n$,
 determined by the method of Ref.~\cite{AOsingular} and by the optimized
 RR method of the present work.}
 \label{comparisonHall}
\end{figure}

The PHO basis has been used in the RR calculations for spiked
oscillators by Hall et al.~\cite{spiked}, who derived useful
formulas for the matrix elements of the operator $r^{k}$ in the PHO
basis on condition of $\gamma>-k/2$. In the case of the "antisextic"
oscillator this amounts to $\gamma>3$, and the expressions for
matrix elements $\langle m|{1\over r^6}|n\rangle$ are singular at
the values $\gamma=0,1,2,3$. Hall et al.~\cite{spiked} used the
prescription of fixing nonlinear parameters
($A=\gamma^2-2\gamma+\frac{3}{4}$ and $B=\Omega^2$) so as to
minimize the approximate energy of the considered level
$E_{nl}^{(N)}(\Omega,\gamma)$. In low orders, an algebraic
diagonalization of the RR matrix has allowed them to obtain
analytical approximations to ground-state energy of the "antisextic"
oscillator, but for its precise determination in higher order
calculation, a time consuming procedure of iterative optimization of
nonlinear parameters must have been used by Naser, Hall and
Katatbeh~\cite{AOsingular}. Our scheme requires much less
computational cost, since the optimum values of nonlinear parameters
are determined from the trace condition and further diagonalization
of the RR matrix is performed only once in each order calculation.
The convergence of our method is demonstrated on two quasi-solvable
examples of the "antisextic" oscillator with angular momentum $l=0$,
where ground-states energies are analytically known. The
semilogarithmic plot of relative errors is shown in
Fig.\ref{singulareven2} for $p=2$ case (
$\lambda_{2(0)}=\frac{5}{384}\left[9887 +
        32\sqrt{333778} \cos[
           \frac{1}{3}\arctan(\frac{1852389 \sqrt{1001}}{478512623})]\right]\approx
           369.26$,
$E_{00}^{(2)}=6$), and in Fig.\ref{singulareven1} for $p=0$ case
($\lambda_{0(0)}=\frac{9}{128}\approx0.07$, $E_{00}^{(0)}=2$). We
observe that for both large and small values of $\lambda$, the
results of the two-parameter ($\Omega_{opt},\gamma_{opt}$) and the
one-parameter optimization ($\Omega=\omega=1,\gamma_{opt}$) are
nearly the same. We have checked that for other spiked oscillators
the case is similar and the calculations may be simplified, since
only parameter $\gamma$ needs to be optimized. This is in difference
with the results of the approach that utilizes iterative
optimization of ground state energy, presented in Table II of
Ref.~\cite{AOsingular}, where much quicker convergence has been
obtained by optimizing both parameters. However, plotting the
relative errors for exited states determined from diagonalization of
the RR matrix with the values of parameters taken from Table II of
Ref.~\cite{AOsingular} for the case of $\lambda=0.1$ in
Fig.~\ref{comparisonHall}, we observe that the precision of energy
determination in this approach rapidly decreases as the number of
the level increases. On contrary, the results of our approach,
obtained from the RR matrix of the same dimension $N=80$, plotted on
the same figure, indicate a uniformly good precision in a wide range
of energy levels. We conclude that the values of nonlinear
parameters obtained from minimization of the trace of the RR matrix
are appropriate for a precise determination of the whole part of the
spectrum, although this is not necessarily the best possible choice
for a particular level.

In Table~\ref{e2one} and ~\ref{e0one} we show the numerical results
of the RR method with the parameter $\gamma$ optimized through
minimization of the trace for the above-discussed quasi-solvable
cases. One can observe how the values of ground state energy and
various moments of the radial position operator converge to the
exact values with increasing dimension of the RR matrix $N$. The
optimum value of $\gamma$ depends strongly on $\lambda$ and grows
with $N$. In all the cases we studied, the condition
$\gamma_{opt}>3$ is fulfilled, although the smaller $\lambda$ is,
the closer $\gamma_{opt}$  is to the value 3, where the matrix
elements become singular, which explains the heavy worsening of
convergence with decreasing $\lambda$. Nevertheless, we may observe
that for $\lambda\rightarrow 0$ the bound-state energies approach
smoothly those of the radial HO, which is due to using the PHO basis
that ensures the satisfaction of the Dirichlet boundary condition in
the RR calculation.

\begin{table}[h]
  \centering
  \begin{tabular}{|c|c|c|c|c|c|}
    \hline
    N &$\gamma_{opt}$ &$ E_{0}^{(N)}(\gamma_{opt})$ & $x_{0}^{(N)}(\gamma_{opt})$ & $x_{0}^{2(N)}(\gamma_{opt})$ & $x_{0}^{6(N)}(\gamma_{opt})$   \\
    \hline
 20  &14.48&  \underline{6.000}21390368223& \underline{2.845}82121384655& \underline{8.28}291890988045&\underline{737}.684487549365\\
 40  &17.29& \underline{6.00000}223509568& \underline{2.8456}2114647890& \underline{8.2818}6647605726&\underline{737.24}9285331635\\
80  & 20.97& \underline{6.0000000}6213529&  \underline{2.84561}903526830& \underline{8.281855}56086959&\underline{737.240}961117416\\
120  & 23.61& \underline{6.00000000}480818&    \underline{2.84561898}882020& \underline{8.281855}31621157&\underline{737.2408}89729308\\
\hline exact& & 6 & 2.84561898466095 & 8.28185529459909 &
737.240860856683\\
    \hline
  \end{tabular}
   \caption{The approximate energy $E_{0}^{(N)}(\gamma_{opt})$
and the moments of the radius operator $r_{0}^{k(N)}(\gamma_{opt})$
(\mbox{$k=1,2,6$}) determined in the optimized RR method for the
ground-state of the radial spiked oscillator~(\ref{sing}) of the
strength $\lambda_{2(0)} \approx 369.26$.}\label{e2one} \label{min}
\end{table}

\begin{table}[h]
  \centering
  \begin{tabular}{|c|c|c|c|c|c|}
    \hline
    N &$\gamma_{opt}$ & $ E_{0}^{(N)}(\gamma_{opt})$ & $r_{0}^{(N)}(\gamma_{opt})$ & $r_{0}^{2(N)}(\gamma_{opt})$ & $r_{0}^{6(N)}(\gamma_{opt})$    \\
    \hline
 20 &4.38 &\underline{2.0}1776622625948& \underline{1.4}4367633236718& \underline{ 2.2}8981024534251& 30.6645428912254\\
 40 & 5.02& \underline{2.00}734163831413& \underline{1.43}695960594035&\underline{2.2}7129896763614& \underline{29}.8195939603567\\
80  &5.88 & \underline{2.00}272386999177 &
\underline{1.43}401033592011& \underline{2.2}6321009528890&
\underline{29}.6572447265661\\ 120  &6.50 &
\underline{2.00}140501517273 & \underline{1.43}316756712602&
\underline{2.2}6090435812050& \underline{29}.7830590372660\\\hline
exact&& 2& 1.43226578557733 & 2.25844053161144& 29.4482015786915\\
   \hline
  \end{tabular}
   \caption{Like Table \ref{e2one}, but for the strength $\lambda_{0(0)}\approx 0.07$.}\label{e0one} \label{min}
\end{table}

\subsubsection{Generalized oscillators}\label{gen}

\begin{figure}[h]
\begin{center}
\includegraphics[width=0.8\textwidth]{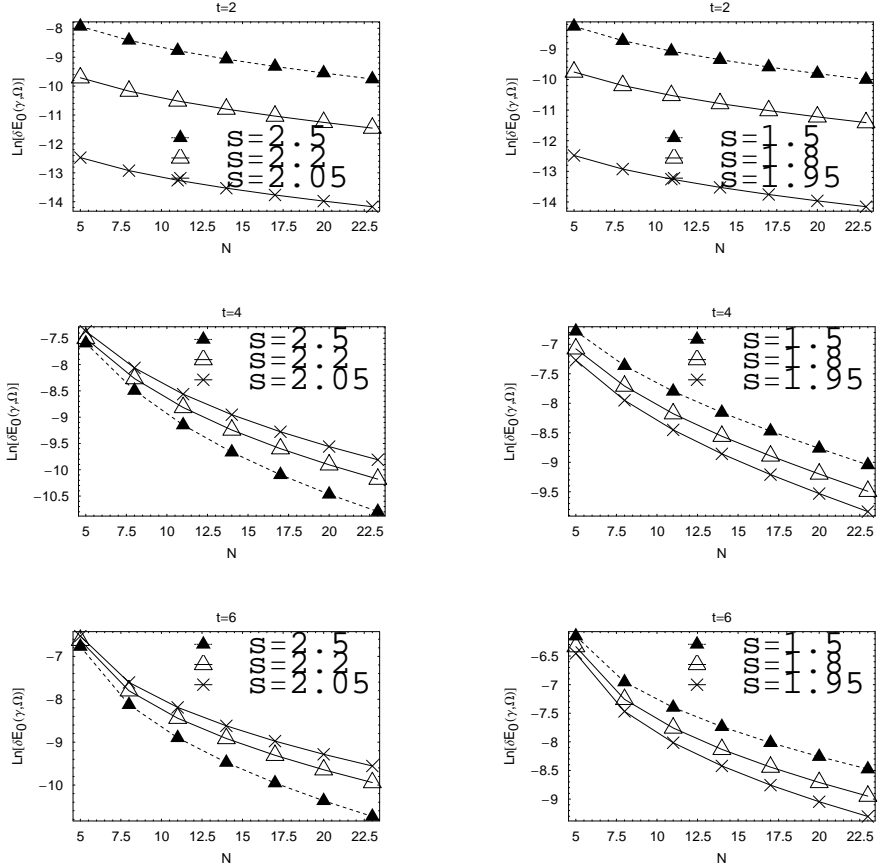}
\end{center}
 \caption{The semilogarithmic plot of the relative error of the optimized
 RR method for bound state
 energies of the generalized oscillator $V(r)=r^s+r^t$ for various
 powers $s$ and $t$,
 as a function of the dimension $N$.}
 \label{generalized}
\end{figure}

We have tested the convergence properties of our approach further by
considering the potential $V(r)$ to be a linear combination of $r^s$
and $r^t$ with the power s being negative and t being positive. The
optimized RR method does allow solution to eigenvalue problem for
various combinations of potential parameters. In
Fig.\ref{generalized} we have plotted the error of ground state
energy determination for the exemplary potential $V(r)=r^s+r^t$ with
various powers $s = -1.5,-1.8,-1.95,-2.05,-2.2,-2.5$ and $t = 2, 4,
6$ in function of the dimension of the optimized RR matrix, $N$. The
convergence becomes exponential at not too large values of $N$, with
the rate depending on the detailed shape of the potential. In all
the cases considered, using the PMS condition for the trace of the
RR matrix for fixing the nonlinear parameters $\Omega$ and $\gamma$
ensures an effective determination of the spectrum.

\section{Conclusion}\label{con}
We have discussed optimization of the RR scheme by introducing
nonlinear parameters, where values are fixed by minimization of the
trace of the truncated matrix. Using the basis of the HO
eigenfunctions with optimized frequency $\Omega$, we obtain an
efficient method for determining spectrum of multi-well
one-dimensional AOs to practically any precision. In the case of
radial oscillators with $\lambda r^{k}$ anharmonicity, the basis of
the PHO eigenfunctions with two arbitrary parameters $\Omega$ and
$\gamma$ seems more suitable. For positive power oscillators ($k>0$)
the role of the parameter $\gamma$ turns out to be minor, and the
scheme may be simplified by using the radial HO eigenfunctions (i.e.
setting $\gamma=3/2+l$). In the case of negative power oscillators
$k<-1$, the parameter $\Omega$ plays a minor role and may be set
equal to the frequency $\omega$ of the harmonic term in the
Hamiltonian, whereas optimization of the parameter $\gamma$ in each
order calculation is crucial for a good convergence. In the limiting
case of harmonium-like potential ($k=-1$) the optimum values of both
parameters turn out to be equal to the ones that are used in the
naive RR method ($\Omega=\omega$ and $\gamma=3/2+l$). The optimized
RR method performs well not only for energies but also for wave
functions, yielding well-convergent approximations to various
moments of the position operator.

The RR method optimized by the trace condition appears effective for
numerical calculation and may be used to arbitrary accuracy within
the modern software environment. It turns out that far greater
improvement of accuracy is obtained from optimizing the nonlinear
parameters in each order calculation by the trace condition rather
than from increasing the dimension of the basis with the parameters
remaining fixed. The computational cost of our scheme is much lower
than in the case of iterative optimization of nonlinear parameters.
Another advantage is that the whole set of energy levels may be
determined at once and the approximate eigenvectors are mutually
orthogonal. For the class of potentials with purely discrete
spectrum considered in the present work, the RR method with the PHO
basis is highly competitive with existing methods, the results of
which are easily recovered in our approach. The convergence of our
approach may be slower than achieved in the methods specialized for
a particular problem, but the advantage is that the results are
obtained automatically for a large class of systems by the algorithm
described above without the necessity of specifying any starting
value.
\appendix
\section{Quasi-exact solutions}
In the appendix we consider the problem of quasi-solvability of
anharmonic oscillators, thereby deriving the formulas for the
analytically known solutions that are used for testing the
convergence of the optimized RR method. Generally, the problem is
quasi-solvable if the eigenfunction can be represented as \be
\psi(y)=f(y)\sum_{n=0}^{p} a_{n}y^{n}\label{series},\ee and the
coefficients of the series satisfy the three-term recurrence
relation  \be A_{n}a_{n+1}+B_{n}a_{n}+C_{n}a_{n-1}=0,~
n=0,1,2,...\label{rec}\ee where $a_{-1}=0$. The coefficients of the
series $a_{n}$ as well those of the recurrence $A_{n}$, $B_{n}$,
$C_{n}$ depend solely on the parameters of the potential and the
bound-state energy $E$. In order for the series in~(\ref{series}) to
terminate after the $p-$th term, we must have \be C_{p+1}=0
\label{C}\ee and \be a_{p+1}=0.\ee It is easy to convince
oneself~\cite{gupta} that the second condition is equivalent to \be
Det \pmatrix{ B_{0}&A_{0}&...&...&0\cr C_{1}&B_{1}&A_{1}&...&0\cr
0&C_{2}&B_{2}&A_{2}&0\cr
...&...&...&...&...\cr0&...&C_{p-1}&B_{p-1}&A_{p-1}\cr
0&...&...&C_{p}&B_{p}\cr}=0,\label{Det}\ee where the determinant is
a polynomial of the degree $p+1$ in the variable $E$. The two
equations (\ref{C}) and (\ref{Det}) may be used to determine the
specific relations between the parameters of the potential that must
be satisfied for the exact solution to be of the form
(\ref{series}).
\subsection{One-dimensional sextic oscillator}
For the one-dimensional sextic AO~(\ref{gen}) the analytically known
eigenfunction, in the even-parity ($\nu=0$) and odd-parity ($\nu=1$)
case, may be represented as \be \psi^{\nu}(x)=e^{-\frac{\lambda
x^4}{2\sqrt{2} }}\sum_{n=0}^{p} a_{n}x^{2n+\nu} \label{states},\ee
where the coefficients of the series satisfy the recurrence
relation~(\ref{rec}) with  \ba A_{n}&=&(2n+2+\nu)(2n+1+\nu),
\nonumber\\B_{n}&=&2E,\nonumber\\
C_{n}&=&-\omega^2-(4n-1 +2\nu)\sqrt{2\lambda}.\ea  The series
in~(\ref{states}) terminates after the $p-$th term, if $C_{p+1}=0$
and $a_{p+1}=0$. The first condition is fulfilled if the parameters
of the sextic oscillator satisfy \be
\omega^2=-(3+4p+2\nu)\sqrt{2\lambda}.\label{condA}\ee  For a chosen
value of $p$, this corresponds to two quasi-solvable cases of a
sextic oscillator: i) $\nu=0$, where the $p+1$ lowest even-parity
states are known; and ii) $\nu=1$, where the $p+1$ lowest odd-parity
states are known. In both cases, we can obtain the $p+1$ exact
bound-state energies $E(\omega,\lambda)$ by solving the polynomial
equation (\ref{Det}), and thus the corresponding exact
wave-functions by determining the non-vanishing coefficients $a_{n}$
from the recurrence relation (\ref{rec}).
\subsection{Radial sextic AO}
For the spherically symmetric sextic AO~(\ref{rad6}) the
analytically known eigenfunction become \be u(r)=r^{l+1}e^{-{\lambda
r^4 \over 2\sqrt{2}}} \sum_{n=0}^{p} a_{n}r^{2n},\ee where the
coefficients $a_{n}$ satisfy the relation (\ref{rec}) with
recurrence coefficients  \ba A_{n}&=&2(n+1)(3+2n+2l),
\nonumber\\B_{n}&=&2E,\label{coef}\\
C_{n}&=&-\omega^2-(1+4n+2l)\sqrt{2\lambda}.\nonumber\ \ea The
closed-form solutions exist if the sextic oscillator parameters
satisfy \be \omega^2=-(5+4p+2l)\sqrt{2\lambda}=0,\label{cond1A}\ee
and the $p+1$ bound-state energies are determined by the condition
(\ref{Det}) with recurrence coefficients of the form (\ref{coef}).

\subsection{Harmonium}
The eigenfunction of the harmonium-like Hamiltonian~(\ref{harg}) can
be written as \be u(r)=r^{l+1}e^{-{\omega\over 2} r^2}\sum_{n=0}^{p}
a_{n}r^{n},\ee where $a_{n}$ satisfy the relation (\ref{rec}) with
recurrence coefficients  \ba A_{n}&=&(n+1)(n+2l+2),
\nonumber\\B_{n}&=&-\lambda,\nonumber\\
C_{n}&=&E-(1+2n+2l)\omega.\label{matrham}\ea The closed-form
solution is obtained if $C_{p+1}=0$, which means that \be E=(3 + 2 l
+ 2 p)\omega,  \label{condH}\ee and the condition (\ref{Det}) is
satisfied with the recurrence coefficients given by (\ref{matrham}).
Compared to the anharmonic oscillator case, the condition
(\ref{condH}) depends on energy; therefore for a particular value of
$\omega$, denoted by $\omega_{p}$, only the bound-state with energy
$E_{p}=(1+2n+2l)\omega_{p}$ is known exactly.

\subsection{Spiked oscillator}
For the spiked AO with the radial Hamiltonian of the form
(\ref{sing}), the analytically known eigenfunction assume the form
 \be u(r)=r^{3\over 2}e^{-{\sqrt{2 \lambda}  \over
2 r^2}-{\omega r^2\over 2}} \sum_{n=0}^{p} a_{n}r^{2n},\ee  where
the coefficients $a_{n}$ satisfy the relation (\ref{rec}) with
recurrence coefficients given by
 \ba A_{n}&=&-4\sqrt{2\lambda}(n+1)
\nonumber\\B_{n}&=&-{3\over 4}-4n(1+n)+l(l+1)+2\sqrt{2\lambda}\omega \nonumber\\
C_{n}&=&-2(E-2n\omega).\label{matrspiked}\ea Closed-form solutions
are obtainable if \be E=2(p+1)\omega,\label{en}\ee and the condition
(\ref{Det}) is satisfied with recurrence coefficients as stated in
(\ref{matrspiked}). For a fixed $p$, Eq.(\ref{Det}) possess $p+1$
solutions that determine the $p+1$ cases of specific relations
between $\lambda$ and $\omega$, at which the bound state of energy
(\ref{en}) is analytically known.

\end{document}